# Historical Auroras in the 990s:
# Evidence for Great Magnetic Storms

Hisashi Hayakawa[1], Harufumi Tamazawa[2], Yurina Uchiyama[3], Yusuke Ebihara[4, 5], Hiroko Miyahara[6], Shunsuke Kosaka[7], Kiyomi Iwahashi[8], Hiroaki Isobe[5, 9]

1. Graduate School of Letters, Kyoto University, 6068501, Kyoto, Japan
   email: hayakawa@kwasan.kyoto-u.ac.jp
2. Kwasan Observatory, Kyoto University, 6078471, Kyoto, Japan
3. Graduate School of Humanities and Sociology, the University of Tokyo, 113-0033, Tokyo, Japan
4. Research Institute for Sustainable Humanosphere, Kyoto University, 6110011, Uji, Japan
5. Unit of Synergetic Studies for Space, Kyoto University, 6068502, Kyoto, Japan
6. Musashino Art University, 1878505, Kodaira, Japan
7. Graduate School of Letters, Tohoku University, 9808576, Sendai, Japan
8. National Institute for Japanese Literature, 1900014, Tachikawa, Japan
9. Graduate School of Advanced Integrated Studies in Human Survivability, Kyoto University, 6068306, Kyoto, Japan

**Abstract** Recently, a significant carbon-14 enhancement in the year 994 in tree rings has been found, suggesting an extremely large cosmic ray flux event during a short period. The origin of this particular cosmic ray event has not been confirmed, but one possibility is that it is of solar origin. Contemporary historical records of low latitude auroras can be used as supporting evidence for intense solar activity around that time. We investigated the previously reported as well as the new records found in contemporary observations from the 990s to determine potential auroras. Records of potential red auroras in the late 992 and early 993 were found around the world, *i.e.* in the Korean Peninsula, Germany, and the Island of Ireland, suggesting the occurrence of an intense geomagnetic storm driven by solar activity.

**1. Introduction: Carbon-14 Variations in 994**

The intensity of a solar flare and its resultant space weather hazards are important issues from both academic and practical points of view. The most intense solar flare and its associated geomagnetic storm recorded by modern telescopic observation was the so-called Carrington event in 1859 (Carrington, 1959; Tsurutani *et al.*, 2003; Hayakawa *et al.*, 2016b; Lakhina and Tsurutani, 2016). The total energy of the Carrington event was estimated to be $\lesssim 10^{32}$ erg (Tsurutani *et al.*, 2003). We have only limited information on the scale of individual solar flares before this event. On the other hand, Maehara *et al.* (2012, 2015) discovered extremely strong "superflares" in solar-type stars (G-type stars with slow rotation), whose estimated energy is $10^{33-35}$ erg. Based on theoretical arguments, Shibata *et al.* (2013) suggested that such superflares may also occur on the Sun. Independently, Miyake *et al.* (2012; 2013) found sharp increases of carbon-14 in tree rings the years from 774 to 775, and from 993 to 994, suggesting significant enhancements in incidental cosmic rays at Earth, which was also supported by the increase of other cosmic-ray induced radioisotopes in ice cores such as beryllium-10 and chlorine-36 (Mekhaldi *et al.*, 2015).

Some studies examined contemporary historical documents to confirm if they provide any evidence for these carbon-14 spikes, as historical documents sometimes involve records of auroras as proxies of solar activity (Vaquero and Vázquez, 2009), up to BCE567 as a literal record and up to 771/772 with drawings (Stephenson *et al.*, 2004; Hayakawa *et al.*, 2016c, d). While the former event in 774/775 became widely discussed (*e.g.* Usoskin *et al.*, 2013; Neuhäuser and Neuhäuser, 2015; Chapman *et al.*, 2015; Hayakawa *et al.*, 2016a), the latter event received far less attention and has only been discussed by Hayakawa *et al.* (2015) and Stephenson (2015). The former, Hayakawa *et al.* (2015), examined Chinese official history and found no relevant records between 988 and 996, but suggested a possibility of records in 992 in the Ulster Annals (*Annala Uladh*) and in German chronicles (Fritz, 1873) to be related with this event. The latter, Stephenson (2015), examined nearby Chinese records and introduced one Korean record from *Goryeosa* (hereafter, GRS)[1] and two German records referring to in the catalogue by Link (1962).

Previous studies detected a cluster of aurora records in the 990s. Fritz (1873) compiled an aurora catalogue from ancient time to 1873 which shows the dates of auroral candidates observed in Germany on 992.04.17, 992.10.21, 992.12.24, 993.01.07[2], and 993.12.26. Link (1962) revisited the

---

[1] Note that Lee *et al.* (2004) and Stephenson (2015) transcribe the same official history (高麗史) as "Koryosa." In our article, we transcribe Korean terms based on the Romanization scheme called "Revised Romanization of Korean."

[2] Although Fritz (1873) misses the date of this event, Short (1749), who is the source for this record, states its date as 992.01.07.

original chronicles and recompiled a new catalogue up to 1600 and integrated them into 992.10.21 and 993.12.26, since the records are quite similar to one another and should have the same origin, except for the one event: 992.10.21. Lee *et al.* (2004) compiled catalogues of auroras and sunspots in Korean historical documents and showed an auroral candidate that could have been observed in 992.12 in Korea. Summarizing, Hayakawa *et al.* (2015) have mentioned a set of records of auroral candidates in 992 in comparison with contemporary Chinese records of auroral candidates. Stephenson (2015) related the German records in Link (1962) and the Korean record in *Goryeosa* for this event.

In this paper, we revisit the contemporary original historical records to analyze their description with original texts and correct some misdating documented in previous studies. We calculate the magnetic latitude of each potential observational site to estimate the Dst scale, and compare reported dates with the sharp spike in carbon-14 reported by Miyake *et al.* (2013).

**2. Method**

In order to reveal the historical evidence of a group of reports of aurora candidates, we restrict this study to the period between 990 and 994 to survey contemporary historical documents all over the world. We introduce and examine relevant historical evidence. The surveyed historical sources and their abbreviations are as follows[3]:

AQ: Anonymous, *Annales Quedlinburgenses*, der SLUB Dresden, Mscr.Dresd.Q.133, Nr.4. [Manuscript in Latin]

ARE-CE: O'Donovan, J. (ed. & trans.) *Annala Rioghachta Eireann: Annals of the kingdom of Ireland by the Four Masters, from the earliest period to the year 1616. Edited from MSS in the Library of the Royal Irish Academy and of Trinity College Dublin with a translation and copious notes*, II. 1856. [Critical edition in Old/Middle Irish]

ARE-MS: Brother Michél Ó Cleitrigh, *Annala Rioghachta Eireann*, Royal Irish Academy, MS1220. [Manuscript in Old/Middle Irish]

AS: Anonymous, *Annalista Saxo*, Bibliothèque nationale de France, MS Latin 11851. [Manuscript in Latin]

AU-CE: Hennessy, W. M. and McCarthy, B. (eds.) *Annala Uladh: Annals of Ulster otherwise

---

[3] In the following reference of historical documents, we consult mainly the original manuscripts. Where we consult both manuscripts and critical editions, we use abbreviations of -MS for manuscripts and -CE for critical editions.

*Annala Senait, Annals of Senat: a chronicle of Irish affairs from A.D. 431 to A.D. 1540.* I., Dublin, 1887. [Critical edition in Old/Middle Irish]

AU-MS: Cathal Óg McMaghnusa, *Annala Uladh*, Trinity College Library, MS1282. [Manuscript in Old/Middle Irish]

CT: Kunstanstalt, F. and Brockmann, O. (eds.) *Die Dresdner Handschrift der Chronik des Thietmar von Merseburg. 1012-1018; Faksimiledruck*, I, 1905. [Manuscript in Latin]

GRS: Jeong Rinji (ed.), *Goryeosa* (高麗史), Waseda University Library, MS ri06_02809. [Manuscript in Chinese]

TYX: *Tiānyuán Yùlì Xiángyìfù*: MS305‑257, Naikaku Bunko, Books of Shoheizaka Gakumonjo, in the National Archives of Japan [Manuscript in Chinese].

We show the relevant records in Section 3. We discuss their historical perspectives, the magnetic latitudes of every observational site, and their relevance to the sharp spike of carbon-14 from 993 to 994.

## 3. Results: Aurora Candidates from 990 to 994

Our survey of aurora records from 990 to 994 brought five records from Saxonian cities in present-day Germany, two records from Ulster in the Island of Ireland, and one record from the Koryo dynasty in the Korean Peninsula. The observational dates are placed on 992.10.21, 992.12.26, and some date between 992.12.27 and 993.01.15. We classify them with respect to their date and observational region. The format for the records is presented as follows: their ID number, their date and place of observation, reference of their original texts[4], their original texts, and our English translation for them.

### 3.1: 992.10.21 in Saxonian cities

(a) AQ: f. 21b

**Original Text:** DCCCCXCII. … XII Calend: Novembris totum cœlom ter in nocte visum est rubrum fuisse.

**Our translation:** 992… On 10.21, the whole sky was seen reddened three times.

---

[4] Although shortened forms are frequently used in the original manuscripts, we show their original form in the original text given in this paper. For example, while "primo" is frequently given as "pmo.," we showed the former to show their original form.

(b) AS: f. 94b

**Original Text:** Anno dccccxcii, … Duodecim Kal. Novembris totum cœlum ter in nocte visum est rubrii fuisse.

**Our translation:** On 10.21, the whole sky was seen reddened three times.

### 3.2 992.12.26 in Saxonian cities

(a) AQ: f. 22a

**Original Text:** DCCCCXCIII. … In nocte natalis sancti protomartyris, id est, VII. Calend. Ianuarii, inauditum seculis miraculum vidimus, videlicet circa primum gallicinium tantam lucem subito ex aquilone effulsisse, ut plurimi dicerent diem oriri. Stetit autem u-nam plenam horam, postea rubente aliquantulum cœlo in solitum conversum est colorem.

**Our Translation:** During the night of St. Protomartyr, *i.e.* on 12.26, we saw an unheard miracle. Around the first cockcrow, suddenly from the North, such a light shined that many people said that the Sun had risen. This continued for a whole hour. Afterwards, the sky was reddened slightly and returned to the normal color.

(b) CT: f. 58a (see, Figure 1)

**Original Text:** In sequenti anno (992) in gallicantu primo lux ut dies ex aquilone effulsit, & unam sic manens horam, undique celo interim rubente, evanuit.

**Our Translation:** In the following year, at the first cockcrow, a light like the sun rose from the North and this continued for one hour. Everywhere in the sky, it was reddened and then vanished.

(c) AS: f. 95b

**Original Text:** Anno dominii vicarii DCCCCXCIII … In nocte natalis sancti Stephani in primo gallicantu lux ut dies ex aquilone effulsit, ut plu-rimi dicerent diem oriri. Stetit autem per unam pleniter horam, et postea rubente aliquantulum celo, in solitum conversum est colorem.

**Our Translation:** On the night of the birth of St. Stephan, at the first cockcrow, light like the Sun shone from the North and many people said the Sun had risen. This continued for a whole hour. Afterwards, the sky was slightly reddened and returned to the normal color.

### 3.3: 992.12.26 in Ulster

(a) AU-MS: f. 53a; AU-CE: p500

**Original Text:** Tadbhsiu ingnadh aidchi feile Stefan, combo croderg in nemh.

**Our Translation:** An unusual appearance on St. Stephan's Eve, and the sky was blood-red.

(b) ARE-MS: f. 388a; ARE-CE: p730

**Original Text:** Fordath teineadh do bith for nimh go matain.

**Our Translation:** The fiery hue was seen in the sky till morning.

**3.3: 992.12-993.1 on the Korean Peninsula**

(a) GRS: 47, f. 19a (see, Figure 2)

**Original Text:** 成宗…十一年十二月夜天門開.

**Our Translation:** In 992.12-993.01[5], at night, heaven's gate opened.

**4. Discussion**

**4.1. Sites and Dates of Observations**

In order to determine the date or site of observation for every record, we examine both the original texts and the background of the historical documents in question.

**4.1.1. Records 3.1: 992.10.21 in Saxonian Cities**

These records were included by Link (1962) and were related to the carbon-14 event in 994 by Stephenson (2015). "Calend" or "Kal." of the month means the first date of the month in question. When this term is placed with dates like "dd C(K)alendas mm[6]" (in 3.1 (a) as "XII Calend: Novembris"), it means "ante diem dd C(K)alendas mm," *i.e.* "dd days before the first date in mm month (including the first date)" (Hampson, 1841), thus "XII Calend: Novembris" or "Duodecim Kal. Nov[embris]" means "12 days before the first date in November" in 992, *i.e.* the date of 10.21 in 992.

The observational sites were not recorded explicitly. However, considering that both AQ and AS were compiled in Saxonian cities, Quedlinburg (N52°47′, E11°09′) and Nienburg (N51°50′, E11°46′), and that the events happened around their hometowns, we expect the observational sites were Saxonian cities around Quedlinburg and Nienburg. The descriptions are quite similar to each other except for small differences; hence, the sources may be the same. Considering that AQ was a contemporary source document compiled in the early 11$^{th}$ century, maybe its text was copied into AS, compiled around 1148-52. Otherwise, we can assume other contemporary chronicles as the common source, such as Annales Hersfeldesnsis or Annales Hildesheimenses maiores (already lost), (Tomaszek, 2010; Dunphy, 2010).

---

[5] Here, only its month is given as "12$^{th}$ month" in traditional calendar while its date is not specified.
[6] Here, dd and mm means date and month.

### 4.1.2. Record 3.2: 992.12.26 in Saxonian Cities

At first sight, these descriptions are similar to each other and we can expect that they may be the same or similar origin, as Link (1962) has pointed out. The dates of (a) and (c) are given as the night of St. Promartyr or St. Stephan, *i.e.* 12.26, although the date of (b) is not written clearly. What is problematic is the year. On one hand, the records (a) and (c) are placed in the sections of the year 993. On the other hand, the author of record (b) places this record "in the following year (in sequenta anno)" of "991 AD (Anno dominicae incarnationis 991)", *i.e.* 992. Link (1962) and Stephenson (2015) have therefore mistakenly judged the date as 993.12.26.

However, we must recall the description of the date in the record (a): "id est VII. Calend. Ianuarii". As we stated above, the term of "Calend. Ianuarii" is placed on the first date of January. Thus, the phrase of "VII. Calend. Ianuarii" is read as "ante diem 7 Calendas Ianuarii", *i.e.* "7 days before the first date in January (including the first date)" (Hampson, 1841). We should subtract seven days from 993.01.01 to recalculate the exact date, considering that record (a) is placed in the first entry of the year 993. We must also note that the beginning of the year was frequently placed on the day of Christmas in Medieval Germany from the 10$^{th}$ to the 15$^{th}$ century after the Carolingian Dynasty (Hampson, 1841: p405; Grotefend, 1898: p12). Therefore, the date in question should be placed not on 993.12.26 as in Link (1962) or Stephenson (2015) but on 992.12.26. This fact can solve the problems of conflicting dates between records (a) and (c) and record (b). The fact that the date given by Calvisius for a quite similar record was 992.12.24 (Calvisius, 1629: p707) supports this solution as well, though he mistakenly places its date on the Christmas Eve.

The observational sites are regarded as Saxonian cities just like record 3.1, as AQ, AS, and CT were all compiled at Saxonian cities, Quedlinburg, Nienburg, and Merseburg (N52°21′, E11°59′). The sources can be the same as the texts are quite similar to each other, even though AQ and CT were compiled in the early 11$^{th}$ century and during 1012-1018, respectively, and hence are regarded as contemporary records (Warner, 2010; Tomaszek, 2010; Dunphy, 2010).

### 4.1.3. Record 3.3: 992.12.26 in Ulster, Ireland

While record (a) of AU is mentioned in Hayakawa *et al.* (2015), record (b) has not been mentioned previously in the context of aurora. The date of St. Stephan is 12.26 (Stokes, 1905). The critical editions and translations of AU (Hennessy and Mc Carthy, 1887; Mc Airt and Mc Niocaill, 1983; CELT, 2000) give its date as "991 alias 992" due to their almanac system. AU places the turn of years up to 1012 not on 1st January but on 25th March, i.e., the Feast of Assumption of St. Mary. Thus, Mc Carthy (1994a) claims to place the date of the AU-MS forward by one year. His idea was accepted by Mc Carthy and Breen (1997) as well. His claim is well confirmed by contemporary records of ARE. The auroral record in ARE is placed in the late section of 992. As the descriptions in AU and ARE are totally different and independent from each other, they are expected to be simultaneous auroral observations by different observers. Considering that ARE places records in chronological order from 1st January to 31st December, we can conclude that the very record of ARE is also placed at the end of the year 992 and indicates a close relationship with the record of AU.

The observational sites were not recorded explicitly. Their texts are totally different from each other, and hence the original observations were most likely carried out independently. AU was compiled by Cathal Óg Mc Maghnusa at Belle Isle on Lough Erne, Ulster (N54°16′, W7°33′) in the late 15th century (Ó Máille, 1910). ARE was compiled by Brother Mícheál Ó Cléirigh from Ballyshannon under the patronage of Fearghal Ó Gadhra at a Franciscan convent of Dunagall (present Donegal; N54°33′ W8°12′) undertaken in 1632 and finished in 1636 (Petrie, 1830). We must note that these annals have excerpts copied from other annals that were previously compiled and lost. Especially, the chapters of AU up to the 9th century were written retrospectively in Middle Irish though the contemporary language in Ireland is Old Irish (Máille, 1910). The chapter for this event is written in contemporary language, i.e., Middle Irish and thus copied from contemporary records. In case these chapters were copied from previous sources such as the Chronicle of Ireland, that is already lost, the observational site can be placed at Armagh (N54°21′, W6°39′), Durrow (N52°51′, W72°24′), or Derry (N55°00′, W7°19′) (Flechner, 2013; Mc Carthy and Breen, 1997; Breeze and Ó Muraíle, 2010).

**4.1.4. Record 3.4: 992.12-993.01 in Gaeseong, Korean Peninsula**

This is the first auroral candidate in Korean historical documents according to Lee et al. (2004). Stephenson (2015) indicates the relationship of this record with the sharp spike of carbon-14 in 994. Its date is placed in the 12th month in the 11th year of Seongjong (成宗), *i.e.* the period between 992.12.27 and 993.01.15 (An *et al.*, 2009).

This record was categorized into a chapter for unusual phenomena on stars including records of red vapor (赤氣) or white vapor (白氣) in the astronomical treatise (天文志) of *Goryeosa* (高麗史). Although the expression of "heaven's gate opened (天門開)" is not so popular for aurora records in East Asian histories, Hayakawa (2016e) examined records including this term to show that they can indicate auroral candidates. The very expression with a similar meaning "heaven's split (天裂)" is frequently found in *Goryeosa*. Similar terminology is used for the first auroral candidate in China in BCE 193 that states "heaven's open (天開)" (*Hànshū*, Astronomy VI: p1710; Saito and Ozawa, 1992; Yau *et al.*, 1995; Hayakawa *et al.*, 2016e) and we can find illustrations for this term in Chinese manuals for astronomical divinations (see Figure 3). We have a Japanese record on 1770.09.17 reporting "heaven's split (天裂)" observed in both Edo and Kanazawa (Hayakawa *et al.*, 2016e), which is contemporary with simultaneous observations of auroras on the same date in China (Willis *et al.*, 1996; Kawamura *et al.*, 2016).

We can assume this was observed at Gaeseong (開城, N37°58′, E126°33′), considering the East Asian tradition to locate observatories near the capital city to report astronomical phenomena to the emperors as soon as possible (Keimatsu, 1970; Hayakawa *et al.*, 2015).

**4.2. Geomagnetic Latitudes of Observational Sites**

Hayakawa *et al.* (2015) showed records of auroral candidates in ancient Chinese chronicles in 960-1279, covering the age of the sharp strike of carbon-14 in 994. In this era, the geomagnetic latitude of East Asia was higher than in the present (Bulter, 1992; Kataoka *et al.*, 2016). Therefore, it was easier to observe the aurora in East Asia (Hayakawa *et al.*, 2015), and a little more difficult to observe it in Europe.

The geomagnetic latitude of observational sites of auroras indicates the relative scale of magnetic storms because the equatorward boundary of the auroral oval is correlated with the Dst index (Yokoyama *et al.*, 1998). The event of 992.10.21 was recorded in Saxonian cities. In the event of 992.12.26, the observation of aurora was recorded simultaneously in Ulster and Saxonian cities. Shortly after this event, a Korean record in GRS follows these records on some date or dates between 992.12.27 to 993.01.15.

Using the global geomagnetic field model CAL3k.4b (Korte and Constable, 2011), we compute the geomagnetic latitude of observational sites in 992-993 and obtain values of approximately 48° for Saxonian cities, 49° for Ulster, and 35° for Gaeseong. Note that the annals in Ulster and Saxonian cities involve the records of nearby areas as stated above. Therefore, an uncertainty of approximately 2° remains as well as the uncertainty of the model. The equatorward boundary of the auroral oval shows a good correlation with the Dst index (Yokoyama *et al.*, 1998). The Dst index is a proxy for the strength of the large-scale electric current, called the ring current, that surrounds the Earth. The prime driver to intensify the ring current is the large-scale magnetospheric convection.

The convection transports hot ions and electrons originating in the plasma sheet on the night side towards the Earth. As they drift earthward due to the convection, the hot ions and electrons gain kinetic energy, and their pressure increases. Because of the spatial difference in the number of ions and electrons, the ring current forms (Ebihara and Ejiri, 2000). Also as the electrons are convected deep into the magnetosphere, they are adiabatically compressed and the electron temperature anisotropy leads to an instability that causes the growth of electromagnetic waves (Kennel and Petschek, 1966; Tsurutani and Lakhina, 1997). The waves scatter the electrons so that they precipitate from the magnetosphere into the upper atmosphere, resulting in aurora. Thus, it is reasonable to consider that the equatorward boundary of the auroral oval is correlated with the earthward boundary of the plasma sheet and the Dst index. Yokoyama et al. (1998) suggested an empirical formula to relate the equatorward boundary of the auroral oval and the Dst index. Assuming that the equatorward boundary of the aurora oval was located at Gaeseong (35° magnetic latitude), we estimate the Dst index to be –970 nT. The minimum value of the Dst index that has been officially published by the World Data Center, Kyoto since 1957 is –589 nT, which was recorded on March 14, 1989. Although the estimated Dst is a crude approximation and a careful diagnosis is necessary, there is a possibility that the magnetic storm that occurred in the period between 992.12.27 and 993.01.15 was a storm larger than any of the storms recorded since 1957.

The magnitude of the magnetic storm is probably not as extreme as that of Carrington event in terms of the equatorward boundary of the auroral oval. During the Carrington event in the equatorward boundaries of the auroral observation were 23° in magnetic latitude, in both the North and South hemispheres (Kimball, 1960; Hayakawa et al., 2016b), and hence its Dst value for this event was estimated by Tsurutani et al. (2003) to be -1760 nT. Uncertainty in estimating the magnitude of the magnetic storm(s) comes from the lack of world-wide historical records of observation, especially in low latitude regions. This may be improved if additional records are discovered. The abnormality of the magnetic storms around the 990s might be explained not by a single extreme magnetic storm, but by several big magnetic storms, as we discuss later.

### 4.3. Long-term Solar Activity

Hayakawa et al. (2015) showed records of auroral candidates in medieval Chinese official histories spanning 960-1279, including the time of the event considered here, in 994. Figure 3 of Hayakawa et al. (2015) showed the peak of the number of auroral candidate observation after the 994 carbon-14 peak (Miyake et al., 2013), before the Oort minimum (1010-1050). The records of auroral observation in the 990s suggest intense solar activity during the period just before the Oort minimum.

We must at the same time consider that Chinese official histories, which had continuous and systematic observations in this age, did not record auroral observations between 988 and 996 (Hayakawa *et al.*, 2015; Stephenson, 2015). We expect that the aurora was simply not observed in this period in China or was not seen due to weather conditions. Otherwise, even if the aurora was observed in this age, it may have been missed in the process of editing official histories from original observational records. The concentration of records of auroral observations may support the occurrence of intense solar activity in these periods, although the relationship with the sharp spike of carbon-14 in 994 is still questionable.

### 4.4. Clustering Observations of Auroras in this Era

One of the features on the records of aurora observations from 992-993 is that we find clustering in the records from 992.12.26 to the period between 992.12.27 and 993.01.15 (Records 3.2, 3.3, and 3.4). Large, flare-productive active regions often generate multiple flares and coronal mass ejections (CMEs), resulting in prolonged space weather disturbances and auroral activities. The Carrington event in 1859 caused a series of auroras lasting as long as 8 days from 08.28 to 09.04 (Loomis, 1859-1865). The auroral observations in 1770.09 provide the earliest known conjugate sighting lasting at least 3 days from 09.16 to 09.18 (Willis *et al.*, 1996; Kawamura *et al.*, 2016). As shown in Tsurutani et al. (2008), the extreme events such as the one in 1972.08 and the Halloween event in 2003.10 are caused by a series of CMEs from a single active region (Mannucci *et al.*, 2005; Tsurutani *et al.*, 2007, Shiota and Kataoka, 2016). Kataoka *et al.* (2016) discussed prolonged auroral activity based on historical documents in East Asia and modern observation. A total of 6 events out of the 20 largest magnetic storms (with a Dst index of < -250 nT) in solar cycles 22 (1986-1996) and 23 (1996-2003) occurred multiply within a week (Kataoka *et al.*, 2016), just like the Halloween event, which featured the clustering of CMEs in October to November 2003 (Mannucci *et al.*, 2005; Tsurutani *et al.*, 2007; Shiota and Kataoka, 2016). Therefore, the auroral observations in late 992 may be regarded as prolonged auroral activity, indicating severe clustering of CMEs in this period.

One may also speculate that the series of CMEs from 992.10 to 993.01 was produced by a single active region. As shown in Tsurutani *et al.* (2008), the extreme events with clustering of CMEs such as the event on 1972.08 with the highest solar wind speed in observational history (Vaisberg and Zastenker, 1976), the above-mentioned Halloween event on 2003.10 with clustered observations of CMEs (Mannucci *et al.*, 2005; Tsurutani *et al.*, 2007), and the Carrington event in 1859 (Carrington, 1859, 1863; Tsurutani *et al.*, 2003) are related with a single active region. It is known that the lifetime of a sunspot $T$ is proportional to the sunspot area $A$: $T \approx A/10$, where $T$ is in days and $A$ is in millionths-of-a-hemisphere (mhs) (the Gnevyshev-Waldmeier law; Gnevyshev, 1938; Waldmeier, 1955; Petrovay and van Driel-Gesztelyi, 1997). According to Notsu *et al.* (2013), the area of the starspots in the superflare stars ranges from a few thousands of mhs, namely comparable to the largest known sunspots, to as large as $10^5$ mhs. If the lifetime of such a huge spot follows the same scaling, the spot should survive months and even years, and may produce flares and CMEs continuously during a significant part of its life. However, the auroral records do not strongly support the recurrency of events with solar rotations. We have not found records of naked-eye sunspots to support the long-lasting intense solar activity reported here.

### 4.5. Relationships with a Sharp Spike of Carbon-14 in 994

Comparison between the records of auroral observations and the sharp spike of carbon-14 in 994 recorded in tree rings (Miyake *et al.*, 2013, 2014) is intriguing. The time lag between carbon-14 input and the absorption by trees has been estimated using the box model of the carbon cycle (see *e.g.* Appendix in Güttler *et al.*, 2015). Güttler *et al.* (2015) discussed the timing of the carbon-14 production that caused the peak of carbon-14 in tree rings in 775, by using a carbon cycle box model as well as the tree-ring records from the northern and southern hemispheres. They suggest that the event had occurred around March 775, with an uncertainty of six months.

According to the case study of the 775 event by Güttler *et al.* (2015), the time lag between the generation of carbon-14 in the atmosphere and the response of isotope ratio in tree rings is several months. Applying the same time scale to the 994 event, under the assumption that the atmospheric circulation is not largely changed, it is suggested that the timing of the event that yielded the carbon-14 spike in 994 should have occurred sometime during late 993 to mid 994. In other words, if the 992.12.26 auroral event was associated with an extremely large cosmic ray flux, it should be reflected as the increase of carbon-14 in 993 as well. We note that such enhancement in 993 is not seen in the carbon-14 data in Japanese trees surveyed by Miyake *et al.* (2013, 2014), but searching for the signature of increased carbon-14 in 993 from other trees, especially those from the southern hemisphere, deserves to be done, to further the discussion on the relationship between the carbon-14 spike in 994 and the historical records of auroras during late 992 to early 993.

### 5. Conclusion

We investigate contemporary aurora records during 990-994. There are five records from Saxonian cities in Germany, two from Ulster in Ireland, and one from the Korean Peninsula. These records

cluster from late 992 to early 993, *i.e.* on 992.10.21, 992.12.26, and some date(s) between 992.12.27 and 993.01.15[7], and show the strong solar activity in at least 992-993 (just before the Oort minimum). The estimated Dst index is −970nT, suggesting that the storm was more extreme than any of the storms formally recorded since 1957 (*e.g.* Mannucci *et al.*, 2005; Tsurutani *et al.*, 2007). The timing of the events is compared with the sharp spikes in carbon-14 shown by Miyake *et al.* (2013). However, the relation is not obvious. We encourage further surveys of records of low latitude auroras and tree-rings in the southern hemisphere in the early 990s.

**Acknowledgement**

We thank Dr. T. Nagamoto for bibliographical information on the medieval western calendar system, Prof. H. Arimitsu for technical advices in transliteration of Saxonian manuscripts, Mr. D. P. Cabezas for grammatical advice, Ms. M. Takakuwa and Ms. M. Hayakawa for giving us advices and insights for medieval western history and contemporary documents, and Dr. F. Miyake for interpretation of the carbon-14 spike in 993/994, although the responsibility for the text rests entirely upon the authors. We also acknowledge the support from the Center for the Promotion of Integrated Sciences (CPIS) of SOKENDAI, the Kyoto University's Supporting Program for Interaction-based Initiative Team Studies "Integrated study on human in space" (PI: H. Isobe), the Interdisciplinary Research Idea Contest 2014 by the Center of Promotion Interdisciplinary Education and Research, the "UCHUGAKU" project of the Unit of Synergetic Studies for Space, and the Exploratory and Mission Research Projects of the Research Institute of Sustainable Humanosphere, Kyoto University. This work was also encouraged by Grant-in-Aid from the Ministry of Education, Culture, Sports, Science and Technology of Japan, Grant Number JP15H05816 (PI: S. Yoden), JP15H03732 (PI: Y. Ebihara), and JP15H05815 (PI: Y. Miyoshi).

**Disclosure of Potential Conflicts of Interest**

The authors describe that they have no conflicts of intersts.

**References**

An, Y.S., Sim, G.G., Song, D.J.: 2009, *Chronological table of Calendar in Koryo Dynasty*, Korean Academic Information, Seoul. [in Korean]

---

[7] Note that GRS only shows us the month in the traditional calendar (the 12$^{th}$ month) and does not specify the date of observation, while other records for aurora-like events are given with the exact date.

Breeze, A.Ó., Muraíle, N.: 2010, *Encyclopedia of the Medieval Chronicle*, Brill, Leiden, I: 93-94.

Butler, R.F.: 1992, *Paleomagnetism: Magnetic Domains to Geologic Terranes*, Blackwell Scientific Publications, Oxford.

Calvisius, S.: 1629, *Opus Chronologicum ex Autoritate Sacrae Scripturae ad Motum Luminarium Coelestium Contextum*, Impensis Johannis Thymii Bibliopolæ, Frankfurt.

Carrington, R.C.: 1859, *Monthly Notices of the Royal Astronomical Society*, 20, 13-15. doi: 10.1093/mnras/20.1.13

Carrington, R.C.: 1863, *Observations of the Spots on the Sun from November 9, 1853, to March 24, 1861, made at Redhill*, William & Norgate, London.

CELT: 2000, *Corpus of Electronic Texts, text ID number: T100001A*, University College Cork, Ireland. http://www.ucc.ie/celt. Accessed 01 Dec 2016

Chapman, J., Neuhäuser, D.L., Neuhäuser, R., Csikszentmihalyi, M.: 2015, *Astronomische Nachrichten*, 336 (6): pp530-544. doi: 10.1002/asna.201512193

Dunphy, G.: 2010, Encyclopedia of the Medieval Chronicle, Brill, Leiden, I: 96-97.

Ebihara, Y. Ejiri, M.: 2000, *Journal of Geophysical Research*, 105 (A7), 15843-15859. doi: 10.1029/1999JA900493

Flechner, R.: 2013, *Early Medieval Europe*, 21.4: 422-454. doi: 10.1111/emed.12025

Gonzalez, W.D., Joselyn, J.A., Kamide, Y., Kroehl, H.W., Rostoker, G., Tsurutani, B.T., Vasyliunas, V.M.: 1994, *Journal of Geophysical Research*, 99, A4, 5771-5792. doi: 10.1029/93JA02867

Grotefend, H.: 1898, *Taschenbuch der Zeitrechnung des deutschen Mittelalters und der Neuzeit*, Hahnsche Buchhandlung, Hannover.

Gnevyshev, M. N.: 1938, *Bulletin de l'Observatoire central Poulkovo*. 24, 37.

Güttler, D. et al.: 2015, *Earth and Planetary Science Letters*, 411: pp290-297. doi: 10.1016/j.epsl.2014.11.048

Hampson, R.T.: 1841, *Medii Aevi Kalendarium*, II, Henry Kent Caston and Co., London.

Hayakawa, H., Tamazawa, H., Kawamura, A.D., Isobe, H.: 2015, *Earth, Planets and Space*, 67:82. doi: 10.1186/s40623-015-0250-y

Hayakawa, H., Isobe, H., Kawamura, A.D., Tamazawa, H., Miyahara, H., Kataoka, R.: 2016a, *Publications of Astronomical Society of Japan*, 68 (3): 33. doi: 10.1093/pasj/psw032

Hayakawa, H., Iwahashi, K., Tamazawa, H., et al.: 2016b, *Publications of Astronomical Society of Japan*, first published online, doi: 10.1093/pasj/psw097

Hayakawa, H., Mitsuma, Y., Fujiwara, Y., et al.: 2016c, *Publications of Astronomical Society of Japan*, in press (arxiv: 1610.08690)

Hayakawa, H., Misuma, Y., Ebihara, Y., Kawamura, A.D., Miyahara, H., Tamazawa, H., Isobe, H.: 2016d, *Earth, Planets and Space*, 68:195. doi: 10.1186/s40623-016-0571-5

Hayakawa, H., Kawamura, A.D., Tamazawa, H., et al.: 2016e, *Sokendai Review of Culutural and Social Studies*, 13, submitted.

Kawamura, A.D., Hayakawa, H., Tamazawa, H., Miyahara, H., Isobe, H.: 2016, *Publications of Astronomical Society of Japan*, 68 (5): 79. doi: 10.1093/pasj/psw074

Kennel, C.F., Petschek, H.E.: 1966, *Journal of Geophysical Research*, 71, 1, 1-28. doi: 10.1029/JZ071i001p00001

Kimball, D.S.: 1960, *A study of the aurora of 1859, Scientific Report No. 6*, the University of Alaska, Anchorage.

Korte, M., Constable, C.: 2011, *Physics of the Earth and Planetary Interiors*, 188 (3–4): 247–259. doi: 10.1016/j.pepi.2011.06.017

Lakhina, G.S., Tsurutani, B.T.: 2016, *Geoscience Letters*, 3:5. doi: 10.1186/s40562-016-0037-4

Lee, E.H., Ahn, Y.S., Yang, H.J., Chen, K.Y.: 2004, Solar Physics, 224, (1-2): 373-386. doi: 10.1007/s11207-005-5199-8

Loomis, E.: 1859, *American Journal of Science*, Second Series, 29, 84: 385-408.

Loomis, E.: 1860a, *American Journal of Science*, Second Series, 29, 85: 92-97.

Loomis, E.: 1860b, *American Journal of Science*, Second Series, 29, 86: 249-265.

Loomis, E.: 1860c, *American Journal of Science*, Second Series, 30, 88: 79-94.

Loomis, E.: 1860d, *American Journal of Science*, Second Series, 30, 90: 339-361.

Loomis, E.: 1861a, *American Journal of Science*, Second Series, 32, 94: 71-84.

Loomis, E.: 1861b, *American Journal of Science*, Second Series, 32, 96: 318-331.

Loomis, E.: 1865, *Annual Report of the Smithsonian Institute*: 208-248.

Mannucci, A.J., Tsurutani, B.T., Iijima, B.A., et al.: 2005, *Geophysical Research Letters*, 32, L12S02. doi: 10.1029/2004GL021467

Mc Carthy, D.: 1994a, *Peritia*, 8: 46-79.

Mc Carthy, D.: 2001, *Early Medieval Europe*, 10.3: 323-341. doi: 10.1111/1468-0254.t01-1-00090

Mc Carthy, D., Breen, A.: 1997, *Vistas in Astronomy*, 41.1: 117-138. doi: 10.1016/S0083-6656(96)00052-9

Mekhaldi, F. et al.: 2015, *Nature Communications*, 6, 8611. doi: 10.1038/ncomms9611

Miyake, F., Nagaya, K., Masuda, K., Nakamura, T.: 2012, *Nature*, 486, 240. doi: 10.1038/nature11695

Miyake, F., Masuda, K., Nakamura, T.: 2013, *Nature Communications*. 4, 1748. doi:

10.1038/ncomms2783

Miyake, F., Masuda, K., Hakozaki, M., Nakamura, T., Tokanai, F., Kato, K., Kimura, K., Mitsutani, T.: 2014, *Radiocarbon*, 56 (3): 1189-1194. doi: 10.2458/56.17769

Máille, T.Ó.: 1910, *The language of the Annals of Ulster*. No. 2. University Press, Manchester.

Neuhäuser, R., Neuhäuser, D.L.: 2015, *Astronomische Nachrichten*. 88: 789-812. 10.1002/asna.201412160

Notsu, Y., Shibayama, T., Maehara, H., Notsu, S., Nagao, T., Honda, S., Ishii, T.T., Nogami, D., Shibata, K.: 2013, *Astrophysical Journal*, 771, 2, 127. doi: 10.1088/0004-637X/771/2/127

Petrie, G.: 1830, *The Transactions of the Royal Irish Academy*, George Bonham, Dublin: 381-393.

Petrovay, K., van Driel-Gesztelyi, L.: 1997, *Solar Physics*, 176, 249. doi: 10.1023/A:1004988123265

Saito, K., Ozawa, K.: 1992, Examination of Chinese ancient astronomical records, Yuzankaku, Tokyo [in Japanese]

Shiota, D., Kataoka, R.: 2016, *Space Weather*, 14 (2): 56-75. doi: 10.1002/2015SW001308

Short, T.: 1749, *General Chronological History of the Air, Weather, Seasons, Meteors, etc. in Sundry Places and Different Times*, Longman & Millar, London.

Stephenson, F.R., Willis, D.M., Hallinan, T.J.: 2004, A&G, 45 (6): 6.15-6.17. doi: 10.1046/j.1468-4004.2003.45615.x

Stephenson, F.R.: 2015, Advances in Space Research, 55 (6): 1537-1545. doi: 10.1016/j.asr.2014.12.014

Stokes, W.: 1905, The Martyrology of Oengus the Culdee: Félire Óengusso Céli Dé. Vol. 29. Harrison and Sons.

Tomaszek, M.: 2010, *Encyclopedia of the Medieval Chronicle*, Brill, Leiden, I: 80-81

Tsurutani, B.T., & Lakhina, G.S.: 1997, *Reviews of Geophysics*, 35, 491-501. doi: 10.1029/97RG02200

Tsurutani, B.T., Gonzalez, W.D., Lakhina, G.S., Alex, S.: 2003, *Journal of Geophysical Research*, 108:A7. doi: 10.1029/2002JA009504

Tsurutani, B.T., Verkhoglyadova, O.P., Mannucci, A.J. et al.: 2007, *Journal of Geophysical Research*, 113, A05311. doi: 10.1029/ 2007JA012879

Tsurutani, B.T., Echer, E., Guarnieri, F.L., Kozyra, J.U.: 2008, *Geophysical Research Letters*, 35, L06S05. doi: 10.1029/2007GL031473

Usoskin, I.G.: 2013, *Living Reviews in Solar Physics*, 10:1. doi: 10.12942/lrsp-2008-3

Usoskin, I.G., Kromer, B., Ludlow, F., Beer, J., Friedrich, M., Kovaltsov, G.A., Solanki, S.K.,

Wacker, L.: 2013 Astronomy & Astrophysics, 552, L3. doi: 10.1051/0004-6361/201321080

Vaisberg, O.L., & Zastenker, G.N.: 1976, Space. Sci. Rev., 19, 687. doi: 10.1007/BF00210646

Vaquero, J.M, Vázquez, M.: 2009, The Sun recorded through History, Springer, Berlin.

Waldmeier, M.: 1955, *Ergebnisse und Probleme der Sonnenforschung*, 2. Auflage, Akademische Verlagsgesellschaft, Leipzig.

Warner, D.A.: 2010, Encyclopedia of the Medieval Chronicle, Brill, Leiden, II: 1424-1425.

Willis, D.M., Stephenson, F.R., Singh, J.R.: 1996, Quarterly Journal of the Royal Astronomical Society, 37: pp733-742.

Yau, K.K.C., Stephenson, F.R., Willis, D.M.: 1995, A catalogue of auroral observations from China, Korea and Japan (193 BC–AD 1770), Chilton.

Yokoyama, N., Kamide, Y., Miyaoka, H.: 1998, Annales Geophysicae, 16, 566. doi: 10.1007/s00585-998-0566-z

**Figure 1**: An aurora-like record in 992.12.26 of Record 3.2 (b) (CT: f. 58a).

**Figure 2**: "Open of Heaven's Gate" of Record 3.4 in GRS (stored in Waseda University Library).

Figure 2 is not available in arXiv version due to license policy. Please kindly see the publisher version of Solar Physics.

**Figure 3**: "Heaven's Split" in *Tiānyuán Yùlì Xiángyìfù*, a Chinese manual of astronomical divination (stored in National Archive of Japan).

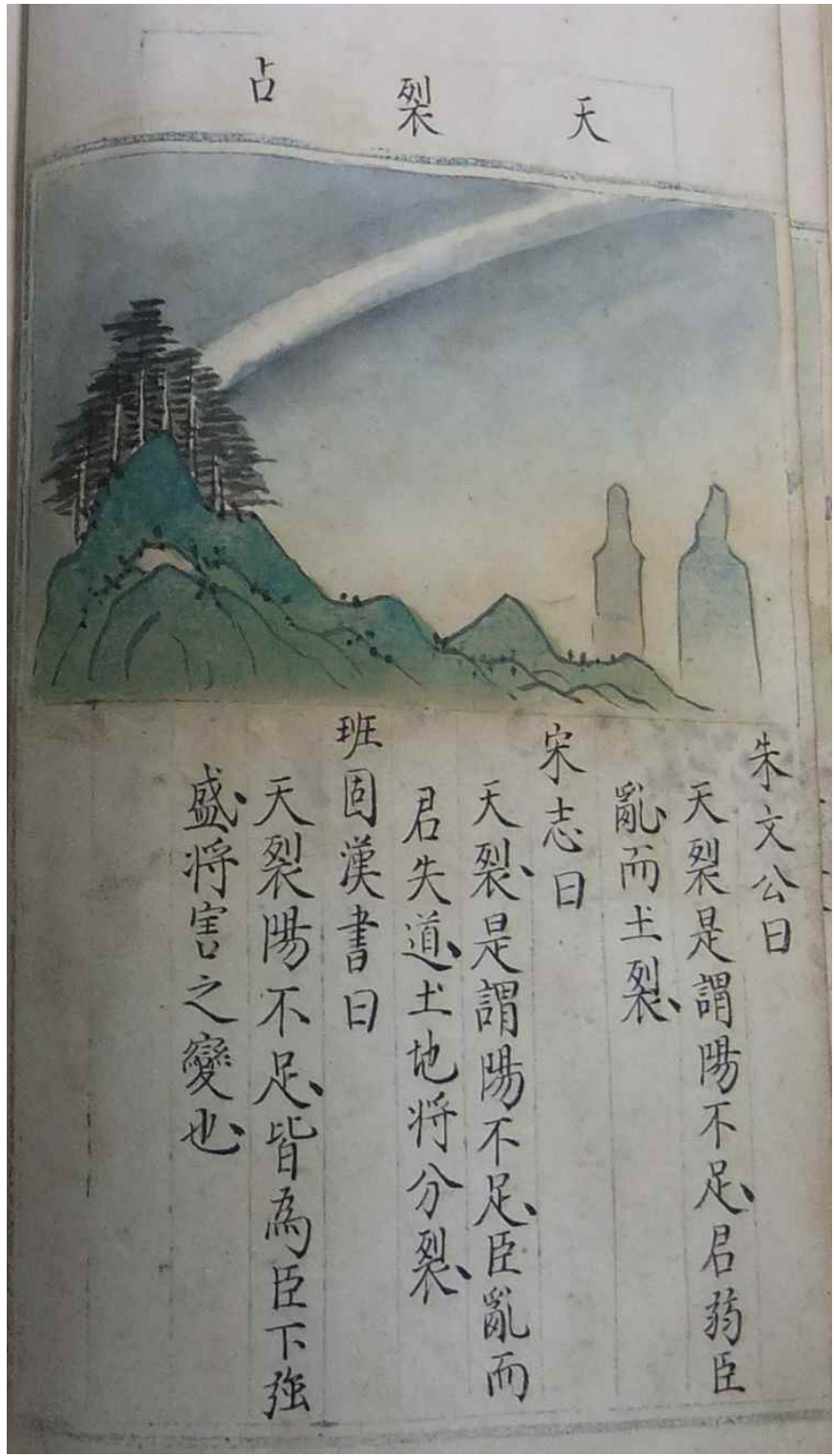